\documentclass[reprint,amsmath,amssym,aps]{revtex4-2}

\usepackage{graphicx}
\usepackage{autobreak}

\newcommand{\apjl}{Astrophys. J. Lett}
\newcommand{\mnras}{Mon. Not. Roy. Astron. Soc.}
\newcommand{\physrep}{Phys. Rep.}

\begin{document}


\title{Fermionic dark stars spin slowly}

\author{Shin'ichirou Yoshida}
\email{yoshida@ea.c.u-tokyo.ac.jp}
\affiliation{Department of Earth Science and Astronomy, Graduate School of Arts and Sciences, The University of Tokyo, \\Komaba 3-8-1, Meguro-ku, Tokyo 153-8902, Japan}


\date{\today}

\begin{abstract}
We study r-mode instability of rotating compact objects composed of asymmetric and
self-interacting Fermionic
dark matter (dark stars). It is argued that the instability limit the angular frequency of the stars
less than half of the Keplerian frequency. This may constrain the stars as alternatives to
fast spinning pulsars or rapidly rotating Kerr black holes.
\end{abstract}


\maketitle

\section{Introduction}
Some alternative models have been presented against such standard compact object models
as neutron stars and black holes. One such category is a compact object made of
dark matter particles. Though nature of dark matter is still elusive, the most plausible
candidates are elementary particles beyond the standard model, which interact
very weakly (or do not interact) with the standard model particles. \cite{2012AnP...524..507F}
These dark matter particles may form a bound state by their self-gravity.
One of the suggestions made is that these exotic objects may be an alternative
to black hole candidates in compact X-ray binaries. However, these alternatives
in accreting X-ray binaries is shown to exhibit thermonuclear flashes which
is absent in observed black hole candidates\cite{2004ApJ...606.1112Y}. 

Objects with typical mass of neutron stars and made from dark matter 
are also considered.They are termed as dark stars \cite{2015PhRvD..92f3526K,2017PhRvD..96b3005M}
and possibility of them to be alternatives to neutron stars is discussed.
These dark stars are made of asymmetric dark matter (ADM) with self-interaction \cite{2014PhR...537...91Z}
which is a candidate of cold dark matter and introduced to solve some cosmological
and astrophysical problems 
\cite{2012MNRAS.423.3740V,2013MNRAS.430...81R,2013MNRAS.431L..20Z,2013MNRAS.430..105P}.

Effects of rotation on these alternatives have not been fully investigated. One of the possible
and important effects is appearance of instability related to stellar rotation. For relativistic
stars, Chandrasekhar-Friedman-Schutz instability \cite{1970PhRvL..24..611C, 1978ApJ...222..281F},
where coupling of stellar oscillations with gravitational wave leads to secular instability for rotating
stars, may be powerful enough to limit the stellar rotation speed
by removing its angular momentum through gravitational wave emission.
Especially the r-mode instability \cite{1998ApJ...502..708A, 1999ApJ...510..846A} may be effective
even for slowly rotating stars as far as the viscous damping of the oscillations are sufficiently weak.

In this paper we study the effect of r-mode instability on the rotational frequency of dark stars.
We do not consider Bosonic dark star, since they may not resemble a fluid star, but rather
a gigantic Bose-Einstein condensate, which require a totally different treatment.
A  conventional recipe of molecular viscosity leads to weak shear viscosity of self-interacting ADM
fluid whose damping effect balances the growth of unstable r-modes. The resulting
spin frequencies of stars are considerably smaller than the Keplerian limit. For a model
with typical mass of a neutron star, the balancing frequency may be lower than
those of observed millisecond pulsars. Higher the mass of the model, the frequency
becomes lower. For a stellar mass black hole case, a dark star model may not
serve as an alternative to a rapidly rotating Kerr black hole. 

\section{\label{sec: formulation}Formulation}
\subsection{Assumptions}
We here study non-rotating stellar equilibria composed of baryonic and dark matter
whose mutual interaction is negligible. The baryonic matter is a zero-temperature nuclear
matter. The dark matter is a self-interacting Fermion with a vector
mediator field. 

\subsection{Dark star model}
Since the stars in the present study are assumed to be static and spherically symmetric,
the spacetime allows the Schwarzschild coordinate $(r,\theta,\varphi)$
in which the spacetime metric is written as, 
\begin{equation}
	ds^2  = g_{\mu\nu}dx^\mu dx^\nu = -e^{2\nu}dt^2 + e^{2\lambda}dr^2 + r^2\sin^2\theta d\varphi^2.
\end{equation}
The geometrized unit, e.g., $c=1=G$, is assumed here. The stress-energy tensor
of the dark matter is written as,
\begin{equation}
	T^{\mu\nu} = (\epsilon+p)u^\mu u^\nu + pg^{\mu\nu}.
\end{equation}
The energy density $\epsilon$ and the pressure $p$ are that of dark Fermion whose equation of
state (EOS) is that of degenerate Fermions of mass $m_{_X}$,
\begin{align}
\epsilon &= \frac{m_{_X}^4}{\hbar^3}\left[\xi(x)+\frac{2}{9\pi^3}\frac{\alpha_{_X}}{\hbar}\frac{m_{_X}^2}{m_\phi^2}x^6\right] ;\nonumber\\
\xi(x) &= \frac{1}{8\pi^2}\left[x\sqrt{1+x^2}(2x^2+1)-\ln(x+\sqrt{1+x^2})\right],
\label{eq: DMeos1}
\end{align}
and
\begin{align}
p &= \frac{m_{_X}^4}{\hbar^3}\left[\chi(x)+\frac{2}{9\pi^3}\frac{\alpha_{_X}}{\hbar}\frac{m_{_X}^2}{m_\phi^2}x^6\right] ;\nonumber\\
\chi(x) &= \frac{1}{8\pi^2}\left[x\sqrt{1+x^2}\left(\frac{2}{3}x^2-1\right)+\ln(x+\sqrt{1+x^2})\right],
\label{eq: DMeos2}
\end{align}
Here $x=p_{_X}/m_{_X}$ is dimensionless Fermi momentum of dark Fermion.
The self-interaction of dark Fermions are mediated by a boson with the rest mass $m_\phi$,
whose coupling constant is $\alpha_{_X}$ (see \cite{Maselli2017} and \cite{Kouvaris2012}).
In this paper we focus our interest on the repulsive self-interaction of dark Fermions, thus $\alpha_{_X}>0$.

Assuming the cross section of the self-interaction to be \cite{2014PhR...537...91Z},
\begin{equation}
	\sigma = 5\times 10^{-23}({\rm cm}^2)
	\left(\frac{\alpha_{_X}}{10^{-2}}\right)^2
	\left(\frac{m_{_X}}{10 {\rm GeV}}\right)^2
	\left(\frac{m_\phi}{10{\rm MeV}}\right)^{-4},
	\label{eq: Zurek-crossection}
\end{equation}
and the constraints of cross section to be \cite{2015PhRvD..92f3526K},
\begin{equation}
	0.1 \le \frac{\sigma}{m_{_X}} \le 1\quad({\rm cm}^2{\rm g}^{-1}),
	\label{eq: Kouvaris-Nielsen-constraint}
\end{equation}
we have a constraint.
\begin{align}
	\begin{autobreak}
	\frac{1}{3}({\rm cm}^2{\rm g}^{-1})\le 
	\left(\frac{\alpha_{_X}}{10^{-2}}\right)^2
	\left(\frac{m_{_X}}{100 m_\phi}\right)^4
	\left(\frac{m_{_X}}{1{\rm GeV}}\right)^2 
	\le \frac{10}{3} ({\rm cm}^2{\rm g}^{-1}). 
	\end{autobreak}
	\label{eq: constraint1}
\end{align}

As is easily seen, we have two independent parameters for dark Fermion EOS.
We choose the following two parameters, $\mu$ and $\kappa$
\begin{equation}
	\mu \equiv \left(\frac{m_{_X}}{1 {\rm GeV}}\right),
	\label{eq: def-mu}
\end{equation}
and
\begin{equation}
	\kappa \equiv 
	\left(\frac{m_{_X}}{100m_\phi}\right)^2\frac{\alpha_{_X}}{10^{-2}}.
	\label{eq: def-kappa2}
\end{equation}
Then the constraint (\ref{eq: constraint1}) is rewritten as
\begin{equation}
	\frac{1}{3} \le \mu^{-3}\kappa^2 \le \frac{10}{3}.
	\label{eq: constraint2}
\end{equation}

With the EOS, Tolman-Oppenheimer-Volkoff equations,
\begin{equation}
	\frac{d\nu}{dr} =  \frac{m+4\pi pr^3}{r(r-2m)},
	\label{eq: dnudr}
\end{equation}
where $m(r)=\int_0^r\epsilon 4\pi r'^2 dr'$ and
the hydrostatic balance is,
\begin{equation}
	\frac{dp}{dr} =  -\frac{(\epsilon+p)(m+4\pi pr^3)}{r(r-2m)},
	\label{eq: dpidr}
\end{equation}
are numerically solved to obtain equilibrium stars.

\subsection{r-mode instability}
We adopt the simple estimation of r-mode instability in \cite{1998PhRvL..80.4843L}.
A stellar model is assumed to be spherically symmetric and the lowest-order post-Newtonian approximation
is adopted to evaluate gravitational radiation from r-mode oscillations.

The gravitational wave timescale $\tau_{GW}$ of r-mode oscillation is estimated by the
ratio of luminosity to oscillation energy as,
\begin{align}
	\begin{autobreak}
		\tau_{GW}^{-1} = -\frac{32\pi G}{c^{2\ell+3}}\Omega_K^{2\ell+2}
		\frac{(\ell-1)^{2\ell}}{[(2\ell+1)!!]^2}
		\left(\frac{\ell+2}{\ell+1}\right)^{2\ell+2}
		\times \int_0^R\rho r^{2\ell+2} dr
		\left(\frac{\Omega}{\Omega_K}\right)^{2\ell+2},
	\end{autobreak}
	\label{eq: tGW}
\end{align}
where $\rho$ is mass density, $\ell$ is the order of spherical harmonics, 
$\Omega$ is the rotational angular frequency, and $\Omega_K$ is the Keplerian
frequency of the given mass $M$ and $R$ of a star \cite{1998PhRvL..80.4843L}.
As for the mass density we simply use
the mass energy density $\epsilon$ divided by $c^2$, which is the solution of TOV equation.

As for viscosity of dark Fermion fluid we adopt the simple estimate of molecular viscosity. 
Microscopic origin of shear viscosity is the transport of momentum by dark matter particles.
The shear viscous coefficient $\eta$ is written as (e.g., see \cite{1992pavi.book.....S})
\begin{equation}
	\eta = a \left(\frac{\sigma_{_X}}{m_{_X}}\right)^{-1}\sqrt{\frac{k_{_B}T_{X}}{m_{_X}c^2}}
	= ac \left(\frac{\sigma_{_X}}{m_{_X}}\right)^{-1}\theta_{_X}^{\frac{1}{2}} 
	\label{eq: shear viscosity}
\end{equation}
where $a$ is a factor of order of unity and $k_{_B}$ is Boltzmann constant. $T_{_X}$
is the 'temperature' of dark Fermions characterizes random motions of dark Fermions.
$\theta_{_X}$ is the normalized temperature parameter $\sqrt{k_{_B}T_{_X}/m_{_X}c^2}$. 
From the assumption of our analysis
on degeneracy of dark Fermion, we expect $\theta_{_X}\ll 1$. We also note that
$\sigma_{_X}/m_{_X}$ and $\theta_{_X}$ are degenerate as parameters.
We here re-normalize $\theta_{_X}$ so that the combination $a\sigma_{_X}/m_{_X}$
is reduced to unity.

The damping timescale $\tau_V$ of r-mode is \cite{1998PhRvL..80.4843L}
\begin{equation}
	\tau_V^{-1} = c(\ell-1)(2\ell+1)\int_0^R\eta r^{2\ell} dr \left( \int_0^R \rho r^{2\ell+2} dr\right)^{-1}.
	\label{eq: tV}
\end{equation}

The critical rotational frequency at which growth of r-mode instability
and viscous damping of the mode are in balance is determined by,
\begin{equation}
	\tau_{GW} + \tau_V = 0,
	\label{eq: rmode balance}
\end{equation}
whose solution $\Omega/\Omega_K$ is the stellar rotational frequency at which both
effect balances. A larger value of $\Omega/\Omega_K$ than $\omega_c$
means r-mode instability dominates and makes the star spin down. The equation is concisely written as
\begin{equation}
	\left(\frac{\Omega}{\Omega_K}\right)^{2\ell+2} 
	= \frac{\tilde{\tau}_V^{-1}}{\tilde{\tau}_{GW}^{-1}} \theta_X^{\frac{1}{2}}
	\equiv \omega_c^{2\ell+2}\theta_X^{\frac{1}{2}}
	\label{eq: omegac def}
\end{equation}
where
\begin{equation}
	\tilde{\tau}_{GW}^{-1} = \frac{32\pi G}{c^{2\ell+3}}\Omega_K^{2\ell+2}
	\frac{(\ell-1)^{2\ell}}{[(2\ell+1)!!]^2}
	\left(\frac{\ell+2}{\ell+1}\right)^{2\ell+2}
	\int_0^R\rho r^{2\ell+2} dr
	\label{eq: tGW def}
\end{equation}
and 
\begin{equation}
	\tilde{\tau}_V^{-1} = c(\ell-1)(2\ell+1)\int_0^R r^{2\ell} dr \left( \int_0^R \rho r^{2\ell+2} dr\right)^{-1}.
	\label{eq; tV def}
\end{equation}
Notice that r-mode timescale is written as $(\Omega_K/\Omega)^{2\ell+2}\tilde{\tau}_{GW}$.

\section{Results\label{sec: Results}}

\subsection{Timescale of r-mode instability and critical angular frequency}
\begin{figure}[hptb]
\includegraphics[scale=0.6]{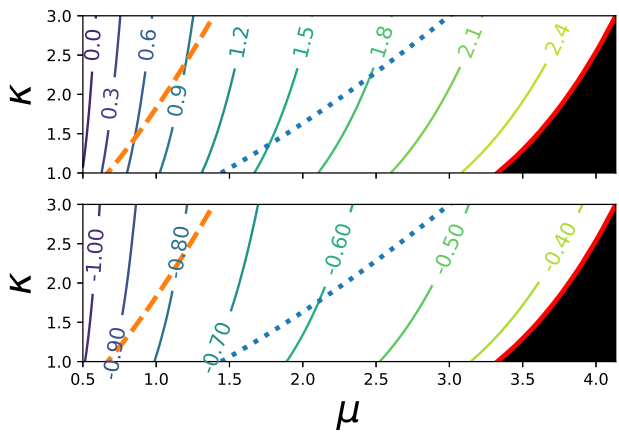}
\caption{Timescale factor $\tilde{\tau}_{_{GW}}$ of r-mode instability and the factor $\omega_c$
of critical angular frequency for $M=1.5M_\odot$ models.
Dark shaded region in the bottom right corner is the excluded region
where the maximum mass of a star is less than $1.5M_\odot$.
Dashed and dotted curves corresponds to the edge of the allowed
parameter region by the current observations (see Eq.(\ref{eq: constraint2})). 
The region between two curves satisfies the constraint. 
Upper panel: Contours of $\log_{10}\tilde{\tau}_{_{GW}}({\rm yrs})$.
Lower panel: Contours of $\log_{10}\omega_c$.
\label{fig: M1.5}
}
\end{figure}

\begin{figure}[hptb]
\includegraphics[scale=0.6]{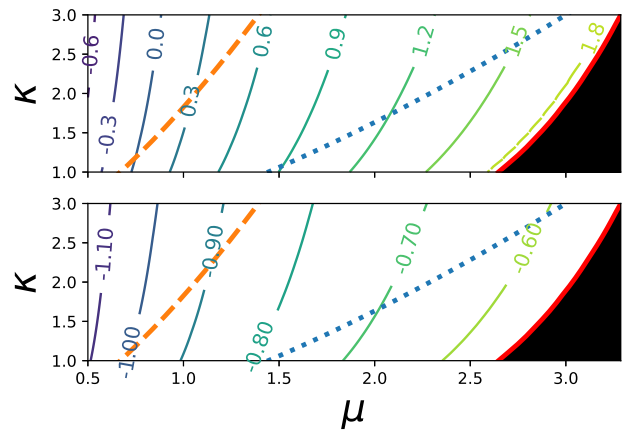}
\caption{Same as Fig.\ref{fig: M1.5} except $M=2.0M_\odot$.
\label{fig: M2.0}
}
\end{figure}

\begin{figure}[hptb]
\includegraphics[scale=0.6]{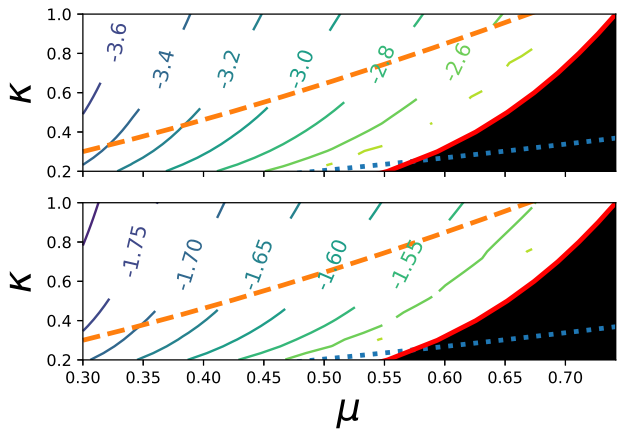}
\caption{Same as Fig.\ref{fig: M1.5} except $M=10M_\odot$.
\label{fig: M10.0}
}
\end{figure}

We compute critical rotation parameter $\omega_c$ and $\tilde{\tau}_{GW}$ as functions
of $\mu$ and $\kappa$ for fixed gravitational mass $M$.  As the Chandrasekhar's limit
of degenerate star is inversely proportional to square of mass of the constituent Fermion,
we have smaller maximum mass for larger $\mu$ when $\kappa$ is fixed. Therefore as
is seen in Fig.\ref{fig: M1.5} the right corners of the parameter space is excluded.
Eq.(\ref{eq: constraint2}) also limit the possible region in the parameters space
consistent with observations (dashed and dotted lines in Figs.\ref{fig: M1.5}, \ref{fig: M2.0},
and \ref{fig: M10.0}), though this may not be a very strong constraint due to 
precision of indirect observations. We simply plot them for references.

In the lower panel of Fig.\ref{fig: M1.5} the contour plot of critical factor $\omega_c$ 
is presented. Gravitational mass $M$ is fixed as $M=1.5M_\odot$. The region between
dashed and dotted curves corresponds to the one satisfying Eq.(\ref{eq: constraint2}).
The factor is ${\cal O}[0.1]$. Therefore the critical angular frequency of the dark star
is at most ${\cal O}[0.1]$, if $\theta_X<1$.
Above this rotational frequency, r-mode instability quickly makes the star spin down.
We conclude that the rotational frequency of dark stars with this mass may be 
smaller than the half of Keplerian frequency.
It is interesting that the fastest known pulsar PSR J1748-2446ad (716Hz) \cite{2006Sci...311.1901H}
may be spinning marginally faster than the limit here, when canonical mass $1.5M_\odot$ 
and radius $12$km are assumed. 
Finding faster pulsars may strongly constrain a dark star as an alternative to millisecond pulsars.

For a reference value, choosing $\mu\sim 1.6$ and $\kappa\sim 2.0$ we have
a ratio of rotational frequency to Keplerian limit $\sim 0.2$ and the time
it takes for a star to spin down from maximal rotation to this value 
is estimated (by using Eq.(\ref{eq: tGW})) to be $2\times 10^5$(yrs).

For the more massive stars as in Fig.\ref{fig: M2.0} and Fig.\ref{fig: M10.0} the trend
is more enhanced. In Fig.\ref{fig: M2.0} the mass is $2.0M_\odot$ which is close
to the theoretical maximum of a neutron star. $\omega_c$ become smaller and 
so does the timescale factor $\tilde{\tau}_{GW}$ since the effect of relativistic 
gravity is stronger.
In Fig.\ref{fig: M10.0} we have $M=10M_\odot$ that corresponds to a typical
stellar mass black hole alternative. For this case the instability timescale is
less than 1yr and the critical rotational frequency is less than 10\% of Keplerian
one. The models have $1-3\times 10^2$km in radius. Therefor the corresponding
dimensionless Kerr parameter is approximately 
$a = 2/5 (\Omega/\Omega_K) \sqrt{GM/c^2R}$, which amounts to 1-2 times
$\Omega/\Omega_K$. It is therefore at most 0.2 for the $M=10M_\odot$ models.
This may be another argument against an attempt to identify a dark star
with a black hole candidate (BHC), since some of the observed stellar-mass BHCs 
in X-ray binaries have the Kerr parameter close to the maximum allowed 
value of unity \cite{1997ApJ...482L.155Z}. 

\section{Summary}
In this paper the effect of r-mode instability in rotating dark stars composed
of asymmetric and self-interacting dark Fermions, which are alternatives
of conventional neutron stars and black holes, are investigated. 
The r-mode instability spins down a star through gravitational wave emission
and viscosity of the dark Fermion fluid tends to damp the instability.
We consider the critical angular frequency for which both mechanisms balances,
which marks the final spin frequency of the star. By applying the conventional
model of molecular viscosity to dark Fermions, we obtain weak shear viscosity.
The critical angular frequency is shown to be less than half of the break-up limit
of the star. The higher the mass, the lower the frequency becomes. For typical
mass of millisecond pulsars ($1.5M_\odot$) to massive neutron stars ($2M_\odot$)
the relative critical frequency to the Keplerian one is less than 20\%, which may
constrain the alternative models in the presence of the fastest spinning pulsars.
For a stellar-mass black hole case ($10M_\odot$), it may be less than 10\%, which may be a
strong constraint to the existence of the alternative model to nearly extreme-Kerr
black holes in X-ray binaries. The timescale of the r-mode instability is also
computed and it is shown to be of astrophysical importance.



\bibliography{dsref}

\end{document}